\documentclass[11pt]{article}
\usepackage{authblk}
\usepackage[titletoc]{appendix}

\usepackage{lineno}

\usepackage{longtable}
\usepackage{booktabs}
\usepackage[symbol]{footmisc}

\usepackage[square,numbers,sort&compress]{natbib}
\usepackage{url}
\usepackage{geometry}
\usepackage{setspace}
\usepackage{rotating}

\usepackage{graphicx}

\title{Fever and mobility data indicate social distancing has reduced incidence of communicable disease in the United States}
\author[1]{Parker Liautaud \footnote{Corresponding author, e-mail: parker\_liautaud@g.harvard.edu}}
\author[1]{Peter Huybers}
\author[2,3]{Mauricio Santillana}
\affil[1]{\small Department of Earth and Planetary Sciences, Harvard University, Cambridge, MA, USA}
\affil[2]{\small Computational Health Informatics Program, Boston Children’s Hospital, Boston, MA, USA}
\affil[3]{\small Department of Pediatrics, Harvard Medical School, Boston, MA, USA}

\begin{document}

\maketitle

\abstract{In March of 2020, many U.S. state governments encouraged or mandated restrictions on social interactions to slow the spread of COVID-19, the disease caused by the novel coronavirus SARS-CoV-2 that has spread to nearly 180 countries. Estimating the effectiveness of these social-distancing strategies is challenging because surveillance of COVID-19 has been limited, with tests generally being prioritized for high-risk or hospitalized cases according to temporally and regionally varying criteria. Here we show that reductions in mobility across U.S. counties with at least 100 confirmed cases of COVID-19 led to reductions in fever incidences, as captured by smart thermometers, after a mean lag of 6.5 days ($90\%$ within 3--10 days) that is consistent with the incubation period of COVID-19. Furthermore, counties with larger decreases in mobility subsequently achieved greater reductions in fevers ($p<0.01$), with the notable exception of New York City and its immediate vicinity. These results indicate that social distancing has reduced the transmission of influenza like illnesses, including COVID 19, and support social distancing as an effective strategy for slowing the spread of COVID-19.}

\section{Introduction}
A new betacoronavirus, SARS-CoV-2, was identified in the city of Wuhan in China's Hubei Province in December, 2019 \citep{wu_2020nature}. Despite efforts to contain it, the disease caused by this virus, COVID-19, spread rapidly across Chinese provinces in January of 2020 and around the world in February and March. As of April 19, 2020, more than 2 million confirmed cases of COVID-19 have led to over 150,000 deaths \citep{total_cases_nyt}, with the United States (U.S.) accounting for a plurality of both.

In the absence of an effective vaccine against the virus or a reliable treatment for COVID-19, the primary public health strategy for slowing transmission has been to encourage or mandate the minimization of all but essential face-to-face human interactions, known as social distancing. Such interventions seek to limit opportunities for transmission by reducing the number of individuals exposed by each newly infected individual. That is, social distancing is intended to decrease the reproductive number of COVID-19 \citep{ma_2020idm, lipsitch2019enhancing}. Implemented social distancing policies in the United States have included closing businesses, restricting large gatherings, and encouraging or requiring residents to stay at home unless undertaking essential tasks \citep{wsj_2020_guide}. These strategies complement other non-pharmaceutical interventions such as contact tracing and isolation or quarantine of individuals suspected to be infected with COVID-19 \citep{cdc_travel_2020, hsiang_2020prp,wsj_2020_guide}. 

Reliable metrics for the impact of social distancing measures on COVID-19 transmission would be useful for informing which are most effective and when and how they should be deployed. Studies of COVID-19 spread in China \citep{chinazzi_2020science, kraemer2020effect,lai2020effect}, Italy \citep{pepe2020covid}, and globally \citep{hsiang_2020prp} have relied upon confirmed case statistics to evaluate the effectiveness of non-pharmaceutical interventions in slowing the spread of the disease.  Confirmed COVID-19 case counts are difficult to interpret, however, owing to uncertainties associated with what fraction of cases are captured in test statistics \citep{Kaashoek_2020,stock2020coronavirus} and high false-negative rates associated with tests \citep{yang2020laboratory}. Furthermore, case reports inevitably lag the onset of illness \citep{wallinga2007generation}. Other studies have used deaths from COVID-19, which are more likely to be recorded than infections, but these represent a small portion of the population and lag disease transmission by two or more weeks \citep{zhou2020clinical}.

Numerous decentralized data sources informing the health and behavior of millions of individuals have become available over the last decade, motivating the development of new methodologies intended to track and ultimately limit the spread of diseases at the population level \citep{salathe2012digital}. To estimate disease activity, these approaches have exploited near real-time data from internet search engines \citep{ginsberg2009detecting, santillana2014can, yang2015accurate}, clinician’s search engines \citep{santillana2014using}, news reports \citep{brownstein2008surveillance}, crowd-sourced participatory disease surveillance systems \citep{smolinski2015flu, paolotti2014web}, Twitter micro-blogs \citep{paul2014twitter}, electronic health records \citep{viboud2014demonstrating, santillana2016cloud}, Wikipedia traffic \citep{generous2014global}, wearable monitoring devices \citep{radin2020harnessing}, satellite images \citep{nsoesie2015monitoring}, and smartphone-based digital thermometers \citep{miller_2018cid}. These new sources of data suggest the potential to achieve near real-time monitoring that is sufficiently complete to infer representative population-wide characteristics and disease trends. In exchange, control over sampling protocols and participant behavior is typically lost, risking the introduction of systematic bias and requiring care in interpretation. 

In this study we use daily estimates of travel distances from cell-phone data and the incidences of fever from a network of smart thermometers across the U.S. to investigate whether the adoption of social distancing practices has slowed the transmission of communicable fever-causing illnesses, including COVID-19. It should be emphasized that the use of novel data to inquire into the dynamics of a novel disease in response to unprecedented changes in social behavior is liable to lead to bias and misinterpretation. By way of example, attempts to predict seasonal influenza from Google search patterns appeared skillful over multiple seasons, but failed to forecast the 2009 H1N1 (``Swine Flu") pandemic and overestimated the spread of the 2013 seasonal influenza in the United States by a factor of nearly two \citep{cook2011assessing,butler_2013nature,lazer2014parable,santillana2014can}. Our aim, therefore, is not to establish the response of COVID-19 to doses of social distancing, but to investigate whether two decentralized data sets may be leveraged to identify patterns consistent with the dynamics of COVID-19.

\section{Data and methods}

Aggregated mobility data are provided by the location analytics company Cuebiq through its Data for Good program. Cuebiq collects first-party location information from smartphone users who opted in to anonymously provide their data through a GDPR-compliant framework. Location data are aggregated into an index, $M$, defined as the base-10 logarithm of median distance traveled per user on each day. For users traversing county borders, their movement is recorded as belonging to the county in which they spend the most time on a given day. In order to better focus our analysis on COVID-19, only counties having at least 100 confirmed cases of COVID-19 as of April 10 are included, giving a total of 368 counties. 

We quantify decreases in mobility between March 5 and April 15 using the range in mobility divided by its maximum value, $\delta M=100 \times (M_{\mbox{max}}-M_{\mbox{min}})/M_{\mbox{max}}$, where the factor of 100 corresponds to reporting $\delta M$ as a percentage. $M$ is first smoothed with a 7-day moving average to eliminate a weekly cycle associated with reduced commuting during weekends.  Although, in principle, $\delta M$ could be ambiguous with respect to increasing or decreasing values of mobility, in practice, all smoothed mobility data show a clear structure of nearly stable values followed by a marked decline. If $\delta M$ was computed between January 1 and March 4 it would average only 10\%, whereas in our selected March 5 to April 15 window $\delta M$ averages 54\%, with 95\% of counties having values between 21\% and 96\%.  These major decreases in mobility coincide with the voluntary and enforced restrictions on movement and business activity introduced in mid-March in many U.S. states \citep{wsj_2020_guide}.

Fever incidence data comes from a network of smart-phone connected personal thermometers managed by Kinsa, Inc. \citep{chamberlain2020real}. Kinsa began monitoring fevers in 2013, and the results of their distributed thermometer network correlate closely with reported ILI cases from the U.S. Centers for Disease Control (CDC) across regions and age groups in previous years \citep{miller_2018cid}.  Aggregated and anonymized fever incidence is reported daily at the level of counties for the contiguous 48 states. Daily fever rates in a given county are calculated as the number of fevers reported divided by the number of unique users within the last year, where a fever is defined as a persistent temperature above $37.7^{\circ}$C.  Despite having more than a million users across the U.S., Kinsa finds it necessary to drop estimates in certain counties whose user base is too small, based upon comparison of their fever tracking against CDC ILI estimates \citep{CDC_ILI}, and instead reports values interpolated from neighboring counties. At the time of writing, we do not know which county estimates are from thermometer data versus being interpolated, and we instead attempt to control for this effect through spatial averaging.  Finally, a linear calibration is applied to the fraction of users experiencing fevers that is determined at the national level in comparison to CDC ILI estimates between August 2016 through August 2019. 

For purposes of analysing whether reductions in mobility are associated with reductions in fevers, we compute a percentage decrease in fevers from Kinsa data similar to that used for mobility, $\delta F = 100 \times (F_{1}-F_{2})/F_{1}$. In this case, $F_1$ and $F_2$ represent the average fever incidence, respectively, over the ten days before a drop in mobility and from ten to twenty days afterwards. The ten-day interval between the end of $F_1$ and beginning of $F_2$ is introduced to account for the time required for a reduction in transmission to manifest as a change in fevers.  The drop in mobility is defined as occurring on the day at which the smoothed mobility signal decreases to halfway between its maximum and minimum values.

\textbf{Limitations.} There are several sources of uncertainty in assessing the degree to which changes in mobility are responsible for changes in fever incidence, foremost that mobility and fever incidence are proxies for the actual variables of interest. The distance traveled by smartphone users  does not, for example, account for the fact that walking on a crowded street involves greater likelihood of close human contact than does driving on an open highway. Similarly, fever data do not distinguish between cases arising from COVID-19, influenza, or any other illness that causes fever. Furthermore, the contribution of COVID-19 is difficult to estimate using fever statistics from previous years because mobility changes are expected to influence transmission of most febrile respiratory infections. The signal of COVID-19 in fever data is also expected to under-represent its actual spread in the population because only approximately 40\% of COVID-19 cases manifest with fevers \citep{nishiura2020estimation}. An additional source of uncertainty arises from seasonality. ILI is strongly seasonal, decreasing through March and April in most years regardless of changes in mobility \citep{moriyama_2020annrev}, and COVID-19 dynamics may also be found to be seasonal, albeit to an unknown extent \citep{NAP25771}. These factors inform our decision to focus on county-level patterns in the rates of fractional fever reduction as opposed to the absolute magnitude of fever declines. 

\section{Results}

Our analysis is divided into three sections. We first examine the relative timing of reductions in mobility and fevers (Figure 1), and next the magnitude of reductions in fevers relative to changes in mobility (Figures 2 and 3). Whereas most counties apparently follow a relationship whereby greater mobility reductions lead to greater reduction in fevers (Figure 3a), counties in the vicinity of New York City follow a distinct pattern (Figure 3b) that we discuss in a final section.

\subsection{Timing of reductions in mobility and fevers}

We first inquire as to the plausibility of a causal relationship between changes in fevers and mobility. Specifically, a mobility-induced change in fever incidence should lag the change of mobility by a time equal to or longer than the incubation period of COVID-19. We estimate the lead or lag between reductions of mobility and reduction in fevers as the time shift that minimizes their mean-squared difference. The lead-lag analysis is computed using time-series of $F$ and $M$ as defined in the methods section, from March 5th, selected for being prior to systematic changes in mobility, to April 15, representing the most current data at the time of our analysis. On average, changes in fevers are found to lag changes in mobility by 6.5 days, with $95\%$ of the 368 counties included in the analysis having lags between 3 and 10 days (Figures 1 and S1). This lag agrees with independent estimates of the incubation period of COVID-19 as averaging 6 days and ranging between 2 and 14 days \citep{lauer2020incubation, liu2020reproductive, li2020early, kucharski2020early}. Observed lags are also consistent with estimates for negative growth rates of COVID-19 cases occurring 8 to 12 days after nation-wide restriction in travel in China \citep{kraemer_2020science}, once accounting for the fact that an additional lag averaging 4.8 days arises between the onset of symptoms, such as fever, and the confirmation of a COVID-19 infection.

Despite consistency with COVID-19 epidemiology, a range of lagged responses to reductions in mobility are expected to be present within our analysis of fevers. Incubation times for various human-to-human communicable and fever-causing diseases range from as short as 1 to 2 days for influenza to several weeks or longer for other diseases \citep{SeattleChildrens}. Although not undertaken herein, more detailed analysis of the relative contributions of COVID-19, influenza, and other illnesses to the fever signal might permit for distinguishing the degree to which close agreement of the observed lag with the COVID-19 incubation period is indicative of COVID-19 prevalence or, instead, arises from a combination of diseases.

\subsection{Magnitudes of reductions in mobility and fevers}

We next consider whether the magnitude of mobility reductions are predictive of the magnitude of the subsequent fever reductions, examining the 368 counties that each have more than 100 confirmed COVID-19 cases. An overall decrease is anticipated solely on the basis of seasonal declines in ILI, but the average seasonally-attributable $\delta F$, as reported by Kinsa, is only expected to average 28\%. In contrast, the observed decline in fevers in 2020 is more than double the expected value with an average $\delta F$ of 79\%. The anomalously large reduction in fevers is consistent with the twin effects of an excess of fevers increasing $F_{1}$ to 31\% above its seasonally-expected value and social distancing subsequently reducing $F_{2}$ to 63\% below its expected value.

Ordinary least-squares regression of $\delta F$ against $\delta M$ indicates a $0.81 \pm 0.28$\% decline in fever incidence with each 10\% reduction in mobility ($p<0.01$, Figure 3a).  Two potential concerns with such a simple regression are that the $p$-value is estimated assuming that counties are independent, and that noise in the predictor values causes regression dilution \cite{frost2000correcting} that biases results toward a shallower slope. A second regression is, therefore, performed after spatially averaging mobility and fever reductions into 5$^\circ$ latitude by 5$^\circ$ longitude bins such that data from 368 counties are represented by 34 grid boxes that are expected to have greater independence and less noise. Regression at this 5$^\circ$ by 5$^\circ$ scale gives a steeper slope indicating a $2.1 \pm 0.6$\% reduction in fever incidence per 10\% reduction in mobility ($p<0.01$) that we consider to be more accurate (Figure S3).

It is useful to briefly review the patterns of mobility and fever reduction across the U.S. that underlie our regression results (Figure 2). The 24 counties in Washington and California with the largest decreases in mobility ($\delta M>90$\%) experienced an average reduction in fever incidences of 90\% (range of 85--93\%). Estimated incidences in these counties all decreased to less then $0.7$\% of the population after mobility reduction, whereas they averaged 4.8\% prior to the change in mobility. More moderate reductions in fever incidence are found in counties with smaller mobility reductions. Counties having $\delta M$ less than 25\%, for example, are all located in South Carolina, Florida, Tennessee, Alabama, Missouri, or Georgia, and are associated with an average reduction in fevers of 80\% (range of 67--89\%).

The fact that the $y$-intercept of our regression of $\delta F$ onto $\delta M$ is $72 \pm 4$\%, using the spatially-averaged data, indicates that there are factors acting to reduce fever incidence nationwide irrespective of changes in mobility. Such a relationship is expected, as noted, on the basis of the seasonality associated with influenza \citep{moriyama_2020annrev}, although seasonality is only expected to account for approximately one-third of the intercept value. A greater-than-expected $y$-intercept may arise if COVID-19 also responds to seasonal changes in environmental conditions, although whether such a sensitivity exists is presently unclear \citep{NAP25771}. Additional factors may include broad adoption of behaviors that limit disease transmission but may not influence travel distances -- such as individuals remaining more dispersed in public, regularly disinfecting surfaces, and more frequently washing their hands -- as well as an approximately 60\% decline in U.S. air travel in March and April relative to 2019 \citep{oag}.

The other major feature associated with the regression is a cluster of outliers that share the distinction of being near New York City (Figure 3a), and which are treated separately in the following subsection. Omitting counties within 500 km of New York City gives a higher county-level regression slope of a $1.6 \pm 0.13$\% decrease in fevers for each 10\% reduction in mobility ($p<0.01$, Figure 3a). At the 5$^\circ$ by 5$^\circ$ degree grid level, again omitting counties within New York City's sphere of influence, the slope is $2.0 \pm 0.4$\% for each 10\% reduction in mobility ($p<0.01$; Figure S3).  Thus, we find a significant regression relationship between mobility and fever reduction whose magnitude  is consistent with the latter estimate of $2.0 \pm 0.4$\% for plausible choices regarding averaging and omission of outliers.

\subsection{New York City and neighboring counties}
As of mid-April, New York City had the highest rate of confirmed COVID-19 infections of any region, accounting for over 30\% of the nation's confirmed cases. Changes in fevers in the five boroughs comprising New York City are, however, not explained by their changes in mobility (Figure 3). Although New York City has achieved amongst the highest levels of mobility reduction, with its five boroughs averaging a $\delta M$ of 81\% and Manhattan having a $\delta M$ of 95\%, the city's fever reductions are amongst the smallest in the country with a $\delta F$ of only 53\%. 

Fever reductions have been more successful further from New York City. A regression of $\delta F$ against distance from Central Park in Manhattan for counties up to 500 km away indicates that $\delta F$ increases by 7\% per additional 100 km distance from Manhattan ($r=0.87$, $p<0.01$). Reductions in mobility, $\delta M$, are greatest in the immediate vicinity of New York City, resulting in a weak regional anti-correlation with $\delta F$ ($r=-0.45$).  County-level population density also weakly corresponds with $\delta F$ such that higher population density is associated with less reduction in fever ($r=-0.38$), as expected \citep{dalziel_2018}. Multiple linear regression using $\delta M$, population density, and distance from New York City to predict $\delta F$, however, indicates that only distance makes a statistically significant contribution.  These results may expose the limitations of our using a single quantification of human mobility as a proxy for human-to-human contact in that it does not capture, for example, if individuals move from more highly infected regions elsewhere. 

As has been found to be generally possible in hierarchical metapopulation models \citep{watts2005multiscale}, we suggest that counties within 500 km of New York City continue to experience relatively high rates of fevers -- despite substantially reducing their local mobility -- because no strict travel restrictions were imposed to protect them from the city's high infection rates.  In contrast with China's restrictions on travel to and from Wuhan that appear to have dramatically reduced the city's ability to seed new outbreaks elsewhere \citep{chinazzi_2020science,kraemer_2020science}, New York City opted instead to focus on local social-distancing efforts \citep{wsj_2020_guide}.  The outstanding question for this region then appears to be why rates of fever have remained relatively high in New York City despite its massive reduction in mobility, with possibilities including that the city is more densely populated than any other in the U.S. \citep{census_pop_2012, dalziel_2018} and that it may have experienced substantial community spread of COVID-19 prior to the implementation of social-distancing rules \citep{NYT20200408}. 

\section{Further discussion and conclusion}
In agreement with foregoing studies \citep{lai2020effect, chinazzi_2020science, hsiang_2020prp, kraemer_2020science}, our results indicate that social distancing has limited the spread of COVID-19. Many U.S. counties with large reductions in mobility achieve near-zero fever incidences by early April, indicating that severe curtailment of mobility is effective in reducing disease transmission. Some counties in Midwestern states, however, achieve similar outcomes with more moderate mobility reductions, and New York City has only partially reduced fevers despite strong decreases in mobility. These counter-examples highlight that regional disparities in demographics, climate, or the delay before implementing mobility restrictions are also potentially of first-order importance for explaining and predicting the spread of COVID-19.

The corollary to social distancing being effective is the question of the optimal degree of mobility restriction for purposes of controlling the spread of COVID-19. Foregoing studies have generally drawn inferences from confirmed case statistics \citep{lai2020effect, chinazzi_2020science, hsiang_2020prp, kraemer_2020science}, but these lag actual infections by up to several weeks and are subject to heterogeneous and potentially biased reporting, making it challenging to infer causality in case growth rates and possibly allowing outbreaks to spread through communities before a response can be mounted.  Our combining cell-phone mobility insights with smart-thermometer fever data permits for more rapidly assessing the timing and rates of change in illness incidence. By leveraging existing data sources for disease surveillance, approaches like the one we have employed may enable timely monitoring of the effectiveness of mobility restrictions and, thereby, support efforts to  contain emerging outbreaks \citep{drake2005limits}.  The importance of such capabilities is underscored by the possibility that COVID-19 outbreaks will recur seasonally and require prolonged or intermittent social distancing \citep{kissler2020projecting}.

\section*{Acknowledgements}
Cuebiq made their county-level mobility data available for this study, and Kinsa made their county-level smart thermometer data public. Comments by Col. Downing Lu (U.S. Army) and R. Scott Kemp (MIT) improved this manuscript. PL and PH were partially supported by the Harvard Global Institute. MS was partially supported by the National Institute Of General Medical Sciences of the National Institutes of Health under Award Number R01GM130668. The content is solely the responsibility of the authors and does not necessarily represent the official views of the National Institutes of Health.

\begin{sidewaysfigure}[ht]
\centering
\includegraphics[width=\linewidth]{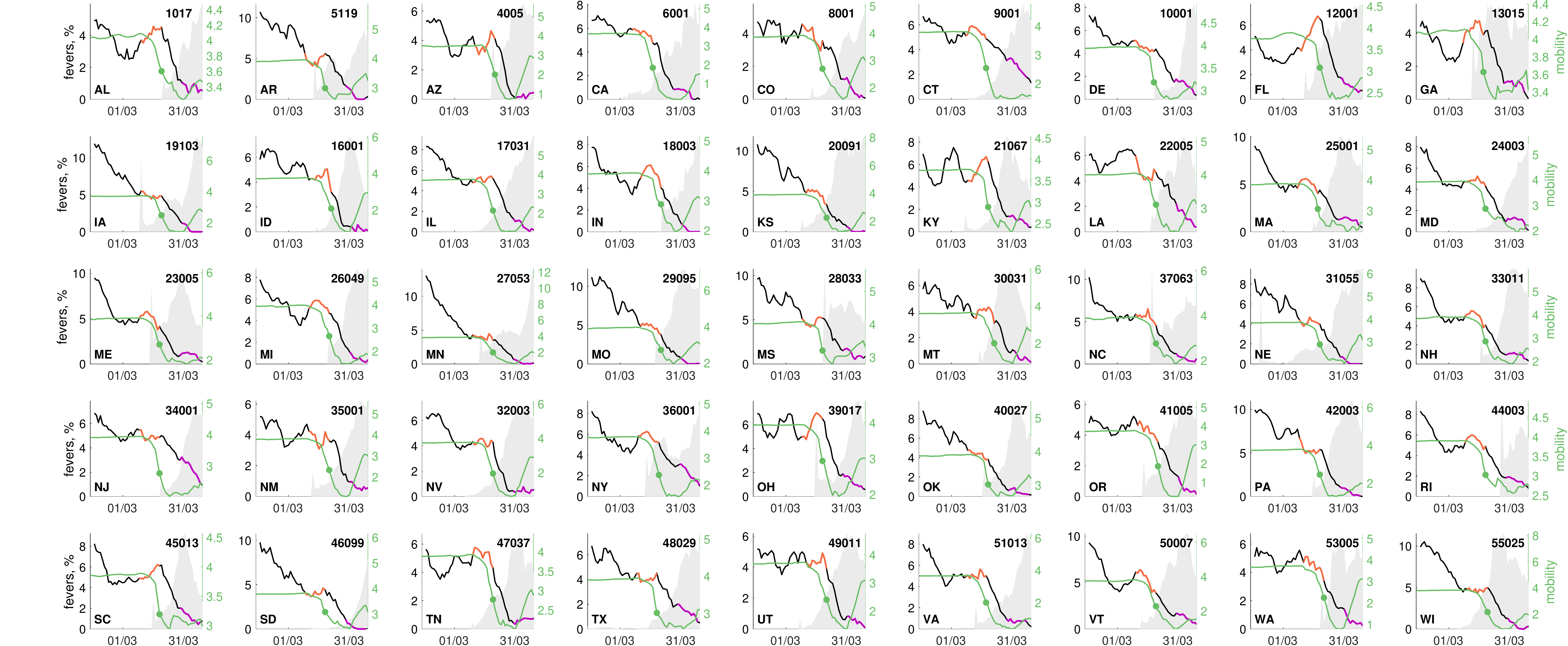}
\caption{Variations in mobility, fever incidence, and confirmed COVID-19 cases for an example county in each of 45 U.S. states. Each panel shows a mobility index (green), the estimated percentage of individuals experiencing fevers (black), and new daily COVID-19 cases (gray shading). The midpoint of mobility reductions (green dot) is used to define an early fever interval, corresponding to $F_1$, over the preceding 10 days (red line), and a later fever interval, $F_2$, between 10 to 20 days afterwards (purple line). COVID-19 cases are normalized such that their maximum value aligns with the upper limit of the y-axis, and the mobility index is centered such that its maximum value equals the average fever incidence during $F_1$. Both COVID-19 cases and mobility are smoothed using a seven-day moving average. Panel labels indicate two-letter state abbreviations and county FIPS code (see Table S1). Omitted states have no county with more than 100 confirmed COVID-19 cases as of April 10, 2020.}
\label{Fig:fig1}
\end{sidewaysfigure}

\begin{figure}[ht]
\centering
\includegraphics[width=\linewidth]{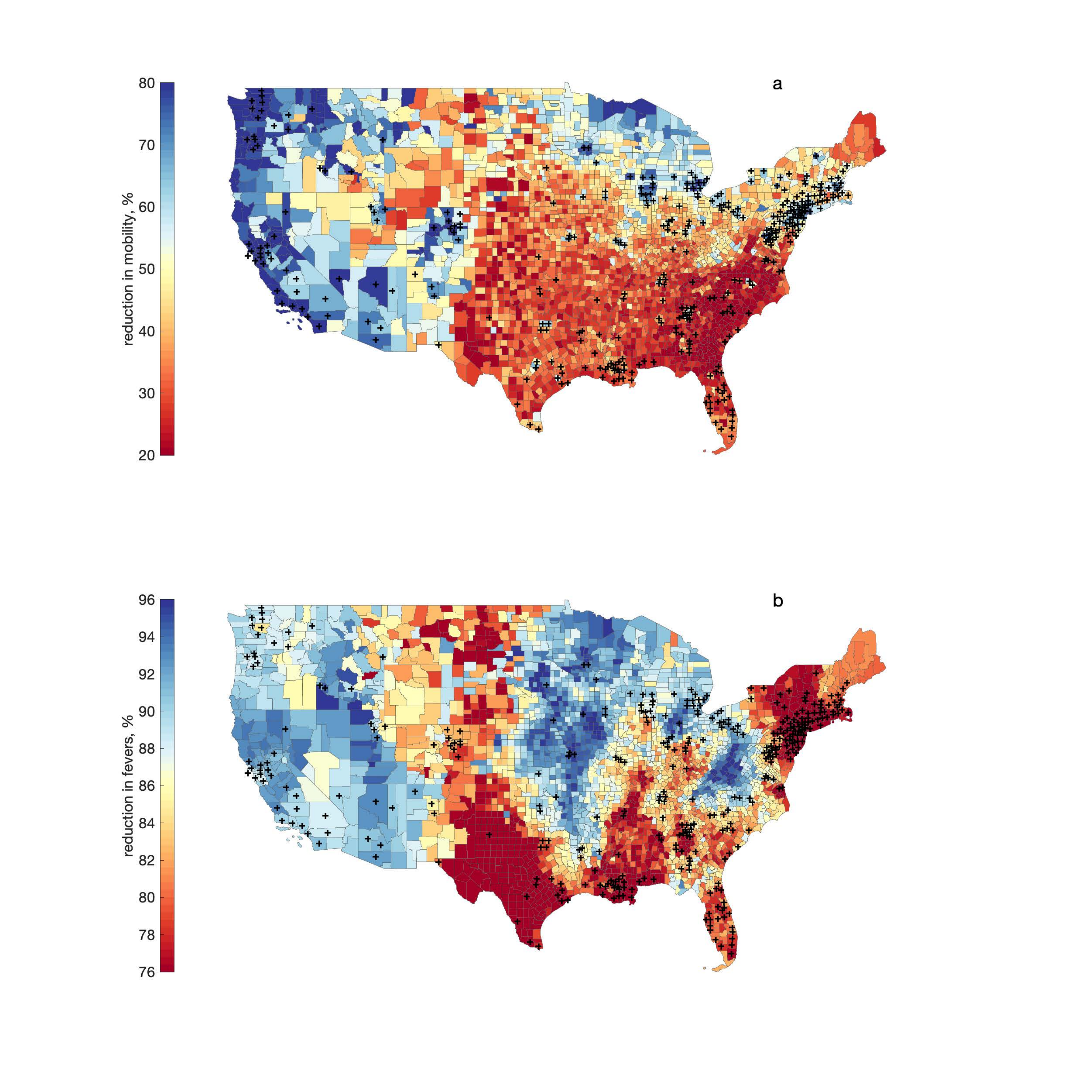}
\caption{Average reductions in mobility and fevers for U.S. counties. \textbf{(a)} Fractional reductions in mobility, $\delta M$, are computed as the range in mobility divided by maximum mobility between March 5th and April 15th, $\delta M=(M_{\mbox{max}}-M_{\mbox{min}})/M_{\mbox{max}}$ \textbf{(b)} Fractional reduction in fever, $\delta F$, computed as the change in fevers across the drop in mobility divided by fevers prior to the drop in mobility, $\delta F = (F_{1}-F_{2})/F_{1}$ (see Figure 1). Counties having more than 100 confirmed COVID-19 cases as of April 10, 2020 (indicated by cross marks) are included in regression analyses shown in Figure 3.}
\label{Fig:fig2}
\end{figure}

\begin{sidewaysfigure}[ht]
\centering
\includegraphics[width=\linewidth]{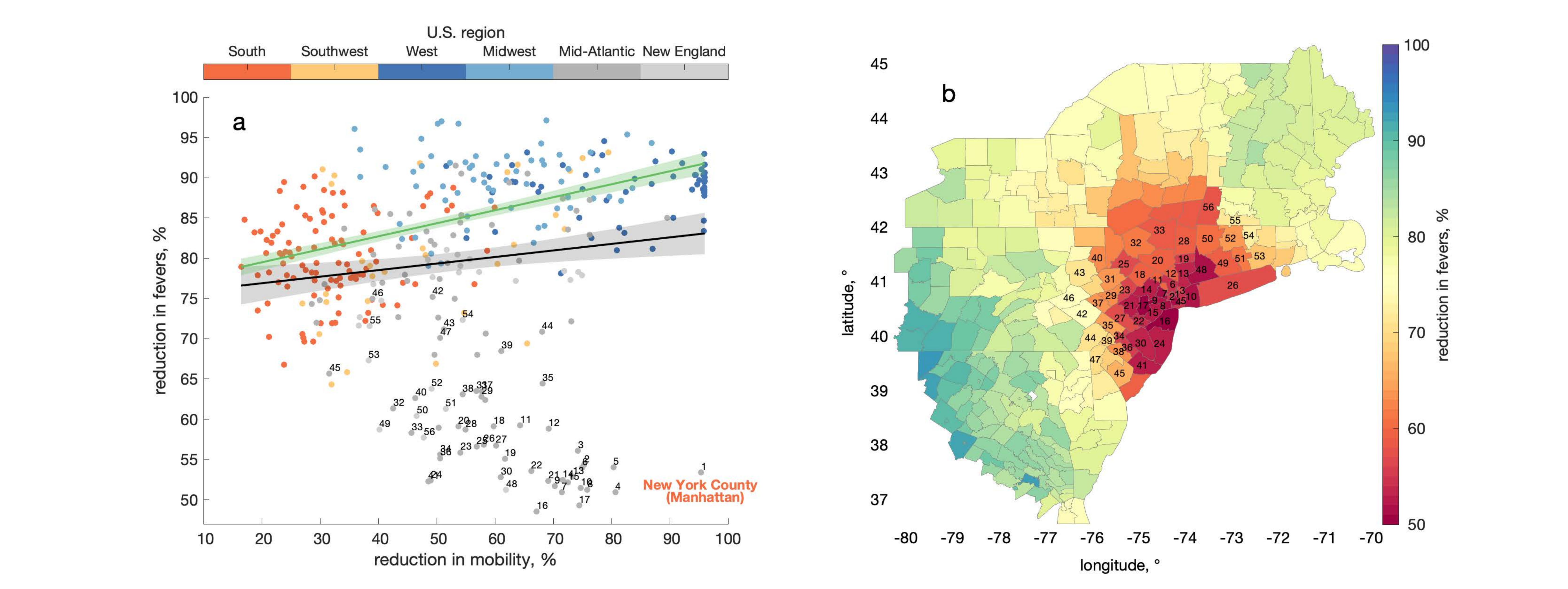}
\caption{Relationship between reductions in mobility and fever across the 368 U.S. counties having more then 100 confirmed COVID-19 cases. \textbf{(a)} Regression of reductions in fevers, $\delta F$, against reductions in mobility, $\delta M$, gives a slope of $0.81 \pm 0.28$\% reduction in fevers per 10\% reduction in mobility (black line and shading). Excluding New York City and counties within 500 km leads to a steeper regression relationship of $1.6 \pm 0.13$\% reduction in fevers per 10\% reduction in mobility (green line and shading). \textbf{(b)} Fractional reduction in fevers, $\delta F$, in counties within 500 km of New York City. Numbering is provided for counties within 200 km of New York City and corresponds to that in panel (a).}
\label{Fig:fig3}
\end{sidewaysfigure}
\clearpage

\clearpage
\bibliography{refs}

\clearpage
\resetlinenumber[1]
\setcounter{page}{1}

\Large{\noindent Supplemental Material for}

\vspace*{5mm}
\Large{\noindent Fever and mobility data indicate social distancing has reduced incidence of communicable disease in the United States}

\vspace*{5mm}
\Large{\noindent Parker Liautaud\footnote[1]{Corresponding author, e-mail: parker\_liautaud@g.harvard.edu}, Peter Huybers, and Mauricio Santillana}

\clearpage
\renewcommand\thetable{S\arabic{table}}  
\setcounter{table}{0}    

\renewcommand\thefigure{S\arabic{figure}}  
\setcounter{figure}{0}   

\begin{figure}[ht]
\centering
\includegraphics[width=0.7\linewidth]{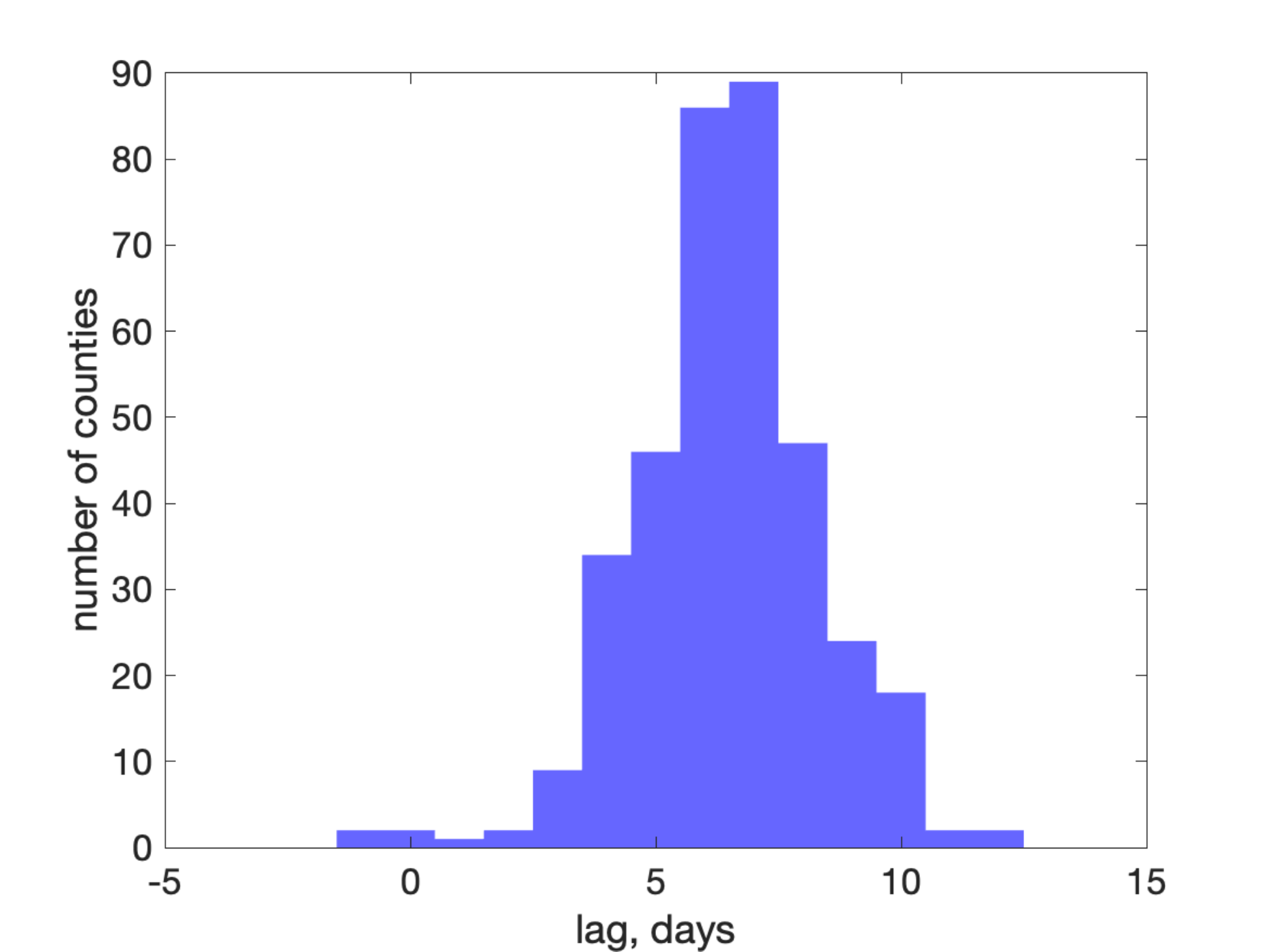}
\caption{Lag of the change in fever incidences, $F$, behind a change in mobility, $M$, for 368 U.S. counties having at least 100 confirmed cases of COVID-19 as of April 10, 2020, where the lag is estimated using data for $F$ and $M$ between March 5 and April 15. The mean lag is 6.5 days with 95\% of counties giving lags between 3 and 10 days. }
\label{Fig:figs1}
\end{figure}

\clearpage
\begin{figure}[ht]
\centering
\includegraphics[width=0.7\linewidth]{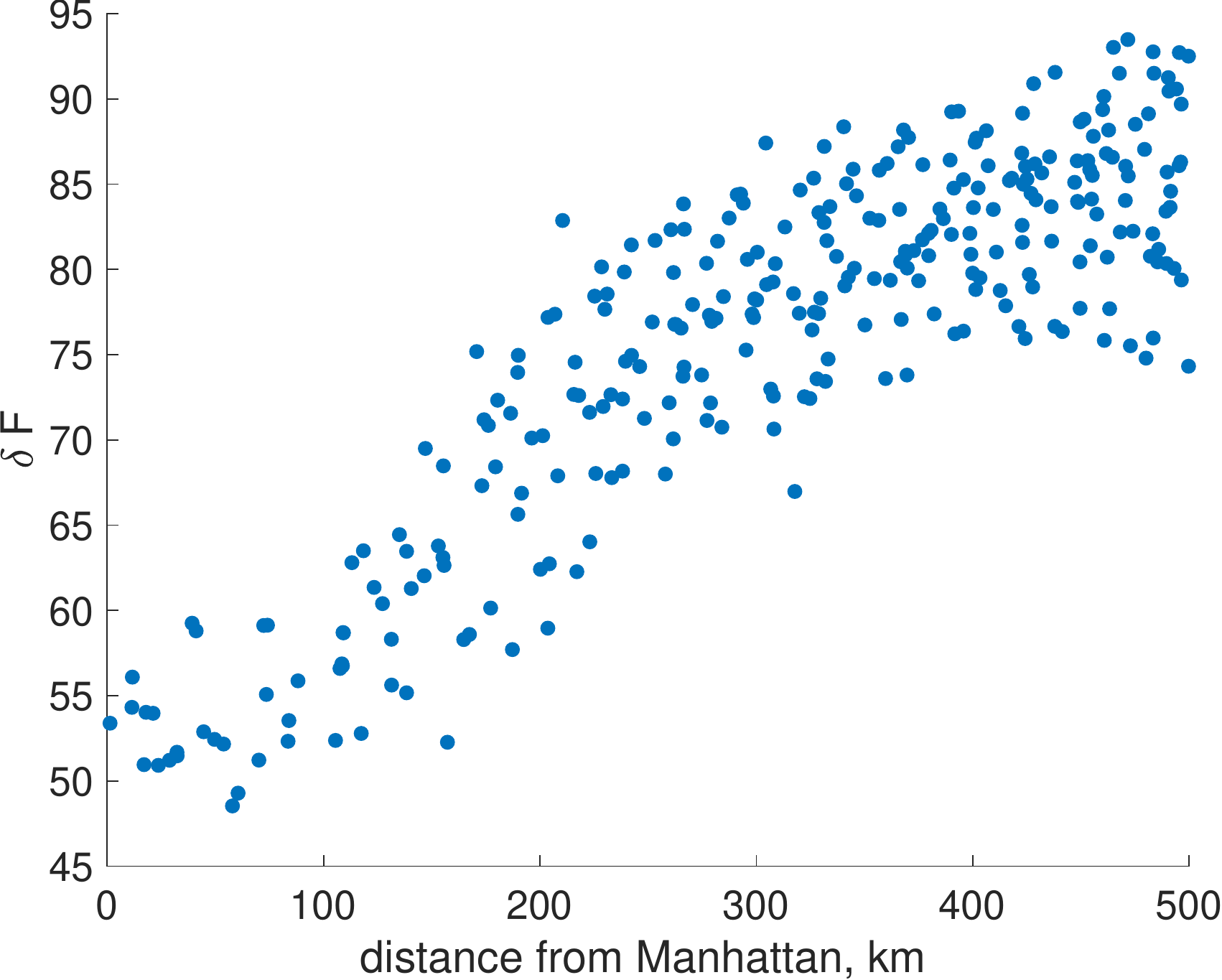}
\caption{Reduction in fever incidences, $\delta F$, for 299 counties that are centered less than 500 km from Manhattan in New York City. Within this region, a county's distance from New York City is strongly correlated with the extent to which its fevers are reduced over a thirty day period ($r=0.87$, $p<0.01$).}
\label{Fig:figs2}
\end{figure}

\clearpage
\begin{figure}[ht]
\centering
\includegraphics[width=0.7\linewidth]{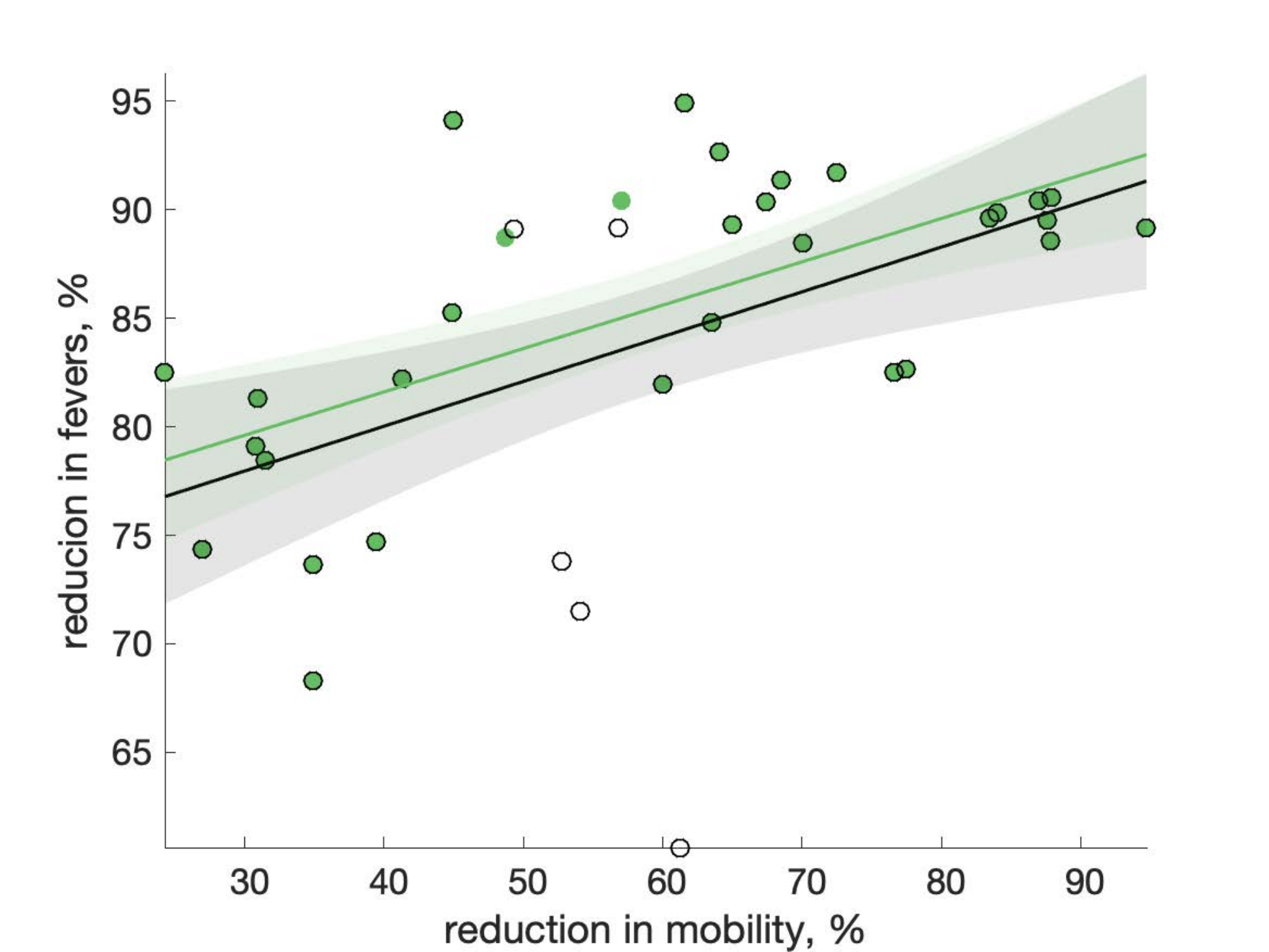}
\caption{Relationship between reduction in fevers and reduction in mobility after spatially averaging counties having more than 100 confirmed COVID-19 cases into 5$^\circ$ latitude by 5$^\circ$ longitude bins. The regression indicates a $2.1 \pm 0.6$\% reduction in fevers for every 10\% reduction in mobility (black markers, line, and shading). If excluding counties within 500 km of New York City, the regression slope is $2.0 \pm  0.4$\% reduction in fevers for every 10\% reduction in mobility (green markers, line, and shading) and the $r^{2}$ increases from 0.27 to 0.42.}
\label{Fig:figs2}
\end{figure}

\clearpage
\begin{spacing}{.8}
\begin{longtable}{llllll}
\caption{Fractional changes in mobility and subsequent changes in fever incidences for the 368 counties shown in Figure ~3 of the main text. Counties are ranked from greatest to least reduction in mobility, and are identified by their Federal Information Processing Standard (FIPS) code.}\\
\label{table:table1}

Rank & County & State & FIPS & $\delta M$, \% & $\delta F$, \% \\
\hline
\endfirsthead

Rank & County & State & FIPS & $\delta M$, \% & $\delta F$, \% \\
\hline
\endhead
1 & Orange County & CA & 6059 & 96.0 & 90.6\\
2 & Ventura County & CA & 6111 & 96.0 & 89.6\\
3 & Placer County & CA & 6061 & 96.0 & 91.6\\
4 & Los Angeles County & CA & 6037 & 96.0 & 90.1\\
5 & Contra Costa County & CA & 6013 & 96.0 & 89.5\\
6 & Snohomish County & WA & 53061 & 95.9 & 88.3\\
7 & Santa Clara County & CA & 6085 & 95.9 & 87.8\\
8 & Kitsap County & WA & 53035 & 95.9 & 88.5\\
9 & San Mateo County & CA & 6081 & 95.9 & 88.1\\
10 & Sonoma County & CA & 6097 & 95.9 & 92.9\\
11 & Washington County & OR & 41067 & 95.9 & 88.3\\
12 & Marin County & CA & 6041 & 95.9 & 88.8\\
13 & Alameda County & CA & 6001 & 95.9 & 88.4\\
14 & Boulder County & CO & 8013 & 95.9 & 83.4\\
15 & Multnomah County & OR & 41051 & 95.8 & 88.5\\
16 & Whatcom County & WA & 53073 & 95.8 & 88.2\\
17 & King County & WA & 53033 & 95.8 & 84.7\\
18 & Island County & WA & 53029 & 95.7 & 90.3\\
19 & San Francisco County & CA & 6075 & 95.7 & 88.6\\
20 & San Diego County & CA & 6073 & 95.4 & 91.0\\
21 & New York County & NY & 36061 & 95.4 & 53.4\\
22 & Skagit County & WA & 53057 & 95.0 & 89.7\\
23 & Spokane County & WA & 53063 & 94.7 & 89.2\\
24 & Clark County & WA & 53011 & 94.5 & 90.3\\
25 & Santa Barbara County & CA & 6083 & 94.3 & 90.3\\
26 & Clackamas County & OR & 41005 & 93.8 & 88.5\\
27 & Sacramento County & CA & 6067 & 92.1 & 91.8\\
28 & Denver County & CO & 8031 & 90.6 & 85.0\\
29 & Pierce County & WA & 53053 & 90.6 & 90.1\\
30 & San Luis Obispo County & CA & 6079 & 90.4 & 93.3\\
31 & Arlington County & VA & 51013 & 89.5 & 85.0\\
32 & Solano County & CA & 6095 & 87.6 & 92.0\\
33 & Douglas County & CO & 8035 & 87.1 & 81.1\\
34 & Blaine County & ID & 16013 & 86.9 & 89.2\\
35 & Hennepin County & MN & 27053 & 86.9 & 94.3\\
36 & Clark County & NV & 32003 & 84.0 & 89.9\\
37 & Washtenaw County & MI & 26161 & 83.6 & 88.7\\
38 & Ramsey County & MN & 27123 & 82.7 & 95.4\\
39 & Marion County & OR & 41047 & 82.4 & 88.2\\
40 & Jefferson County & CO & 8059 & 82.2 & 83.1\\
41 & Oakland County & MI & 26125 & 81.6 & 87.1\\
42 & Eagle County & CO & 8037 & 81.3 & 83.8\\
43 & Arapahoe County & CO & 8005 & 80.9 & 81.0\\
44 & Kings County & NY & 36047 & 80.7 & 51.0\\
45 & Queens County & NY & 36081 & 80.4 & 54.0\\
46 & DuPage County & IL & 17043 & 79.6 & 87.6\\
47 & Coconino County & AZ & 4005 & 79.5 & 93.2\\
48 & Benton County & WA & 53005 & 79.1 & 87.2\\
49 & Ada County & ID & 16001 & 78.7 & 94.5\\
50 & Washoe County & NV & 32031 & 78.4 & 92.8\\
51 & Dane County & WI & 55025 & 77.6 & 90.9\\
52 & Lake County & IL & 17097 & 76.9 & 87.4\\
53 & Fresno County & CA & 6019 & 76.7 & 93.0\\
54 & Larimer County & CO & 8069 & 76.6 & 82.5\\
55 & Ingham County & MI & 26065 & 76.4 & 93.2\\
56 & Montgomery County & MD & 24031 & 76.3 & 87.2\\
57 & Fairfax County & VA & 51059 & 76.2 & 82.9\\
58 & Richmond County & NY & 36085 & 75.9 & 51.2\\
59 & Gallatin County & MT & 30031 & 75.6 & 90.3\\
60 & Hudson County & NJ & 34017 & 75.3 & 54.3\\
61 & Pima County & AZ & 4019 & 75.1 & 92.4\\
62 & Riverside County & CA & 6065 & 75.0 & 88.6\\
63 & Chittenden County & VT & 50007 & 75.0 & 77.9\\
64 & Bergen County & NJ & 34003 & 75.0 & 54.0\\
65 & Nassau County & NY & 36059 & 74.7 & 51.5\\
66 & Summit County & UT & 49043 & 74.5 & 85.0\\
67 & Somerset County & NJ & 34035 & 74.5 & 49.3\\
68 & Suffolk County & MA & 25025 & 74.3 & 78.3\\
69 & Bronx County & NY & 36005 & 74.2 & 56.1\\
70 & Alexandria County & VA & 51510 & 74.0 & 85.9\\
71 & Westchester County & NY & 36119 & 73.4 & 52.9\\
72 & Tompkins County & NY & 36109 & 73.1 & 72.2\\
73 & Norfolk County & MA & 25021 & 73.0 & 77.3\\
74 & Kane County & IL & 17089 & 73.0 & 86.3\\
75 & Maricopa County & AZ & 4013 & 72.9 & 90.8\\
76 & Milwaukee County & WI & 55079 & 72.6 & 91.1\\
77 & Cook County & IL & 17031 & 72.6 & 87.6\\
78 & Loudoun County & VA & 51107 & 72.6 & 83.5\\
79 & Middlesex County & NJ & 34023 & 72.5 & 52.2\\
80 & Bernalillo County & NM & 35001 & 72.0 & 83.6\\
81 & Middlesex County & MA & 25017 & 71.9 & 78.4\\
82 & San Joaquin County & CA & 6077 & 71.6 & 91.5\\
83 & Howard County & MD & 24027 & 71.6 & 87.4\\
84 & Morris County & NJ & 34027 & 71.6 & 52.4\\
85 & Essex County & NJ & 34013 & 71.5 & 50.9\\
86 & Waukesha County & WI & 55133 & 71.4 & 89.5\\
87 & El Paso County & CO & 8041 & 71.3 & 78.1\\
88 & Macomb County & MI & 26099 & 71.1 & 92.8\\
89 & McHenry County & IL & 17111 & 70.7 & 86.2\\
90 & Sandoval County & NM & 35043 & 70.4 & 85.4\\
91 & Union County & NJ & 34039 & 70.3 & 51.7\\
92 & Rockland County & NY & 36087 & 69.2 & 58.8\\
93 & Hunterdon County & NJ & 34019 & 69.2 & 52.3\\
94 & Johnson County & KS & 20091 & 68.8 & 97.1\\
95 & St. Louis County & MO & 29510 & 68.5 & 84.0\\
96 & Livingston County & MI & 26093 & 68.4 & 91.6\\
97 & Montgomery County & PA & 42091 & 68.2 & 64.5\\
98 & Wayne County & MI & 26163 & 68.1 & 92.1\\
99 & Chester County & PA & 42029 & 68.1 & 70.9\\
100 & Grant County & WA & 53025 & 68.0 & 88.9\\
101 & Olmsted County & MN & 27109 & 67.9 & 94.2\\
102 & St. Louis County & MO & 29189 & 67.4 & 83.8\\
103 & San Bernardino County & CA & 6071 & 67.4 & 88.2\\
104 & Monmouth County & NJ & 34025 & 67.2 & 48.5\\
105 & Will County & IL & 17197 & 66.7 & 81.9\\
106 & Salt Lake County & UT & 49035 & 66.5 & 89.2\\
107 & Mercer County & NJ & 34021 & 66.2 & 53.5\\
108 & Hamilton County & IN & 18057 & 66.2 & 92.1\\
109 & Allegheny County & PA & 42003 & 65.6 & 90.5\\
110 & Utah County & UT & 49049 & 65.5 & 93.1\\
111 & Travis County & TX & 48453 & 65.5 & 69.4\\
112 & Kent County & MI & 26081 & 65.4 & 89.2\\
113 & Passaic County & NJ & 34031 & 64.3 & 59.3\\
114 & Monroe County & NY & 36055 & 64.1 & 79.7\\
115 & Pinal County & AZ & 4021 & 63.9 & 91.6\\
116 & Genesee County & MI & 26049 & 63.7 & 90.8\\
117 & Hampshire County & MA & 25015 & 63.6 & 77.2\\
118 & Davis County & UT & 49011 & 63.5 & 89.4\\
119 & Collin County & TX & 48085 & 63.2 & 81.7\\
120 & Kern County & CA & 6029 & 63.0 & 89.5\\
121 & Stanislaus County & CA & 6099 & 62.9 & 91.5\\
122 & St. Charles County & MO & 29183 & 62.8 & 86.2\\
123 & Butler County & PA & 42019 & 62.4 & 89.4\\
124 & Navajo County & AZ & 4017 & 62.1 & 90.7\\
125 & Johnson County & IA & 19103 & 62.1 & 92.7\\
126 & Cuyahoga County & OH & 39035 & 62.0 & 89.7\\
127 & Fairfield County & CT & 9001 & 61.9 & 51.2\\
128 & Putnam County & NY & 36079 & 61.7 & 55.1\\
129 & Delaware County & PA & 42045 & 61.1 & 68.5\\
130 & Burlington County & NJ & 34005 & 61.0 & 52.8\\
131 & Denton County & TX & 48121 & 60.5 & 79.3\\
132 & Lorain County & OH & 39093 & 60.3 & 91.7\\
133 & Bucks County & PA & 42017 & 60.2 & 56.7\\
134 & Adams County & CO & 8001 & 60.1 & 83.3\\
135 & Williamson County & TN & 47187 & 59.9 & 86.3\\
136 & Sussex County & NJ & 34037 & 59.8 & 59.1\\
137 & Franklin County & OH & 39049 & 59.7 & 88.4\\
138 & Yakima County & WA & 53077 & 59.5 & 88.2\\
139 & Erie County & NY & 36029 & 59.5 & 84.1\\
140 & Kenosha County & WI & 55059 & 59.3 & 87.1\\
141 & Tulare County & CA & 6107 & 59.1 & 90.1\\
142 & Lake County & OH & 39085 & 58.9 & 88.7\\
143 & Fulton County & GA & 13121 & 58.6 & 80.9\\
144 & Onondaga County & NY & 36067 & 58.5 & 70.6\\
145 & Albany County & NY & 36001 & 58.4 & 62.4\\
146 & Summit County & OH & 39153 & 58.2 & 89.4\\
147 & Saginaw County & MI & 26145 & 58.1 & 90.4\\
148 & Suffolk County & NY & 36103 & 58.1 & 56.9\\
149 & Lehigh County & PA & 42077 & 57.7 & 63.5\\
150 & Northampton County & PA & 42095 & 57.7 & 62.8\\
151 & Essex County & MA & 25009 & 57.3 & 78.3\\
152 & Saratoga County & NY & 36091 & 56.9 & 68.0\\
153 & Pike County & PA & 42103 & 56.9 & 56.6\\
154 & Monroe County & PA & 42089 & 56.8 & 63.5\\
155 & Lucas County & OH & 39095 & 56.7 & 92.7\\
156 & Orleans County & LA & 22071 & 56.5 & 76.8\\
157 & Portage County & OH & 39133 & 56.3 & 89.8\\
158 & St. Clair County & MI & 26147 & 56.1 & 91.5\\
159 & Monroe County & MI & 26115 & 55.7 & 91.2\\
160 & Worcester County & MA & 25027 & 55.6 & 79.9\\
161 & St. Clair County & IL & 17163 & 55.3 & 85.2\\
162 & Hamilton County & OH & 39061 & 55.2 & 83.5\\
163 & Weld County & CO & 8123 & 55.2 & 82.5\\
164 & Jackson County & MI & 26075 & 55.0 & 91.9\\
165 & Dutchess County & NY & 36027 & 55.0 & 58.7\\
166 & Fort Bend County & TX & 48157 & 54.7 & 73.2\\
167 & Tolland County & CT & 9013 & 54.5 & 72.3\\
168 & Gloucester County & NJ & 34015 & 54.4 & 63.1\\
169 & Frederick County & MD & 24021 & 54.2 & 85.4\\
170 & Beaver County & PA & 42007 & 54.2 & 90.5\\
171 & Carroll County & MD & 24013 & 54.1 & 84.4\\
172 & Warren County & NJ & 34041 & 54.1 & 55.9\\
173 & Rockingham County & NH & 33015 & 54.0 & 79.0\\
174 & Plymouth County & MA & 25023 & 53.8 & 78.2\\
175 & Orange County & NY & 36071 & 53.7 & 59.1\\
176 & Polk County & IA & 19153 & 53.7 & 96.7\\
177 & Cumberland County & ME & 23005 & 53.6 & 80.4\\
178 & Lake County & IN & 18089 & 53.6 & 88.1\\
179 & Westmoreland County & PA & 42129 & 53.2 & 91.5\\
180 & Dauphin County & PA & 42043 & 53.1 & 75.0\\
181 & Durham County & NC & 37063 & 52.9 & 88.8\\
182 & Lancaster County & PA & 42071 & 52.8 & 82.9\\
183 & Wake County & NC & 37183 & 52.6 & 86.5\\
184 & Mahoning County & OH & 39099 & 52.3 & 91.8\\
185 & Niagara County & NY & 36063 & 51.9 & 79.4\\
186 & Forsyth County & GA & 13117 & 51.8 & 86.0\\
187 & Hillsborough County & NH & 33011 & 51.8 & 81.0\\
188 & Canyon County & ID & 16027 & 51.7 & 91.4\\
189 & Middlesex County & CT & 9007 & 51.6 & 61.3\\
190 & Hendricks County & IN & 18063 & 51.5 & 87.2\\
191 & Luzerne County & PA & 42079 & 51.2 & 71.2\\
192 & St. Joseph County & IN & 18141 & 51.1 & 88.9\\
193 & Jackson County & MO & 29095 & 50.8 & 97.0\\
194 & Butler County & OH & 39017 & 50.7 & 79.4\\
195 & New Castle County & DE & 10003 & 50.6 & 70.1\\
196 & Camden County & NJ & 34007 & 50.6 & 55.2\\
197 & Philadelphia County & PA & 42101 & 50.6 & 55.6\\
198 & Anne Arundel County & MD & 24003 & 50.5 & 77.4\\
199 & Cape May County & NJ & 34009 & 50.4 & 59.0\\
200 & Broome County & NY & 36007 & 50.3 & 72.6\\
201 & Allen County & IN & 18003 & 50.2 & 96.7\\
202 & Williamson County & TX & 48491 & 49.9 & 66.9\\
203 & York County & PA & 42133 & 49.6 & 81.7\\
204 & Schenectady County & NY & 36093 & 49.6 & 68.0\\
205 & Franklin County & MA & 25011 & 49.3 & 77.7\\
206 & Berks County & PA & 42011 & 49.3 & 75.2\\
207 & Hartford County & CT & 9003 & 49.1 & 63.8\\
208 & Kankakee County & IL & 17091 & 49.1 & 80.6\\
209 & Linn County & IA & 19113 & 49.1 & 92.7\\
210 & Ocean County & NJ & 34029 & 49.0 & 52.4\\
211 & Prince William County & VA & 51153 & 49.0 & 82.1\\
212 & Davidson County & TN & 47037 & 48.9 & 88.0\\
213 & Decatur County & IN & 18031 & 48.6 & 79.0\\
214 & Atlantic County & NJ & 34001 & 48.6 & 52.3\\
215 & York County & ME & 23031 & 48.5 & 79.8\\
216 & Mecklenburg County & NC & 37119 & 48.4 & 87.6\\
217 & Prince George's County & MD & 24033 & 48.3 & 83.3\\
218 & El Paso County & TX & 48141 & 48.1 & 80.4\\
219 & Berkshire County & MA & 25003 & 47.8 & 57.7\\
220 & DeKalb County & GA & 13089 & 47.7 & 86.5\\
221 & Brazos County & TX & 48041 & 47.6 & 81.2\\
222 & Marion County & IN & 18097 & 47.5 & 91.4\\
223 & Douglas County & NE & 31055 & 47.4 & 94.8\\
224 & San Juan County & NM & 35045 & 47.2 & 91.8\\
225 & Litchfield County & CT & 9005 & 46.6 & 60.4\\
226 & Stark County & OH & 39151 & 46.5 & 88.9\\
227 & St. James County & LA & 22093 & 46.4 & 76.8\\
228 & Lackawanna County & PA & 42069 & 46.4 & 62.6\\
229 & Montgomery County & OH & 39113 & 46.3 & 82.5\\
230 & Trumbull County & OH & 39155 & 46.0 & 90.8\\
231 & Cobb County & GA & 13067 & 45.8 & 82.5\\
232 & Ulster County & NY & 36111 & 45.7 & 58.3\\
233 & James City County & VA & 51095 & 45.4 & 85.5\\
234 & Oneida County & NY & 36065 & 45.0 & 77.2\\
235 & Baltimore County & MD & 24005 & 44.6 & 77.9\\
236 & East Baton Rouge County & LA & 22033 & 44.2 & 76.9\\
237 & Miami County & OH & 39109 & 44.0 & 82.4\\
238 & Johnson County & IN & 18081 & 43.6 & 84.3\\
239 & Harford County & MD & 24025 & 43.5 & 81.4\\
240 & Lebanon County & PA & 42075 & 43.4 & 72.7\\
241 & Ascension County & LA & 22005 & 43.3 & 74.1\\
242 & Sullivan County & NY & 36105 & 42.6 & 61.4\\
243 & Richmond County & VA & 51760 & 42.0 & 85.5\\
244 & Madison County & IN & 18095 & 41.3 & 91.3\\
245 & Dallas County & TX & 48113 & 41.1 & 78.3\\
246 & Shelby County & TN & 47157 & 40.7 & 74.2\\
247 & Baltimore County & MD & 24510 & 40.6 & 77.0\\
248 & Barnstable County & MA & 25001 & 40.6 & 74.7\\
249 & New Haven County & CT & 9009 & 40.2 & 58.7\\
250 & Sumner County & TN & 47165 & 39.9 & 85.6\\
251 & Alachua County & FL & 12001 & 39.7 & 79.6\\
252 & Henrico County & VA & 51087 & 39.4 & 86.1\\
253 & Madison County & AL & 1089 & 39.2 & 86.5\\
254 & Bristol County & MA & 25005 & 39.2 & 76.8\\
255 & Charles County & MD & 24017 & 39.2 & 80.5\\
256 & Cameron County & TX & 48061 & 39.0 & 77.8\\
257 & Schuylkill County & PA & 42107 & 39.0 & 75.0\\
258 & St. Tammany County & LA & 22103 & 38.9 & 74.8\\
259 & Fayette County & KY & 21067 & 38.8 & 81.0\\
260 & Montgomery County & TX & 48339 & 38.6 & 79.1\\
261 & Hampden County & MA & 25013 & 38.5 & 71.6\\
262 & New London County & CT & 9011 & 38.3 & 67.3\\
263 & Shelby County & AL & 1117 & 38.3 & 73.2\\
264 & Hidalgo County & TX & 48215 & 38.1 & 74.7\\
265 & Iberville County & LA & 22047 & 38.0 & 78.7\\
266 & Jefferson County & AL & 1073 & 37.8 & 72.2\\
267 & Jefferson County & KY & 21111 & 37.5 & 83.2\\
268 & Collier County & FL & 12021 & 37.3 & 77.9\\
269 & Tarrant County & TX & 48439 & 37.2 & 79.6\\
270 & Sedgwick County & KS & 20173 & 36.9 & 90.1\\
271 & Minnehaha County & SD & 46099 & 36.9 & 93.5\\
272 & Palm Beach County & FL & 12099 & 36.8 & 78.0\\
273 & Providence County & RI & 44007 & 36.8 & 72.7\\
274 & Kent County & RI & 44003 & 36.7 & 71.6\\
275 & Lee County & FL & 12071 & 36.7 & 80.5\\
276 & Orange County & FL & 12095 & 36.1 & 77.4\\
277 & Wyandotte County & KS & 20209 & 35.9 & 96.1\\
278 & Broward County & FL & 12011 & 35.8 & 77.9\\
279 & Osceola County & FL & 12097 & 35.3 & 77.5\\
280 & Lafayette County & LA & 22055 & 35.1 & 78.4\\
281 & Washington County & LA & 22117 & 34.6 & 79.2\\
282 & Bexar County & TX & 48029 & 34.6 & 65.8\\
283 & Gwinnett County & GA & 13135 & 34.2 & 86.6\\
284 & Chesterfield County & VA & 51041 & 34.2 & 83.6\\
285 & St. Martin County & LA & 22099 & 34.0 & 81.8\\
286 & Pulaski County & AR & 5119 & 33.9 & 88.2\\
287 & Wilson County & TN & 47189 & 33.7 & 82.2\\
288 & Tangipahoa County & LA & 22105 & 33.5 & 76.6\\
289 & Seminole County & FL & 12117 & 33.5 & 77.5\\
290 & Plaquemines County & LA & 22075 & 33.2 & 73.1\\
291 & Miami-Dade County & FL & 12086 & 33.2 & 75.1\\
292 & Sarasota County & FL & 12115 & 33.2 & 79.4\\
293 & St. Johns County & FL & 12109 & 32.8 & 77.2\\
294 & St. Charles County & LA & 22089 & 32.6 & 72.0\\
295 & Cherokee County & GA & 13057 & 32.6 & 81.8\\
296 & Tulsa County & OK & 40143 & 32.5 & 90.3\\
297 & Forsyth County & NC & 37067 & 32.2 & 88.1\\
298 & Lafourche County & LA & 22057 & 32.1 & 71.9\\
299 & Bossier County & LA & 22015 & 32.1 & 85.2\\
300 & Martin County & FL & 12085 & 32.0 & 82.3\\
301 & Webb County & TX & 48479 & 32.0 & 64.3\\
302 & Cleveland County & OK & 40027 & 32.0 & 89.2\\
303 & Hinds County & MS & 28049 & 31.9 & 77.3\\
304 & Cumberland County & NJ & 34011 & 31.6 & 65.6\\
305 & DeSoto County & MS & 28033 & 31.3 & 76.3\\
306 & Harris County & TX & 48201 & 31.1 & 74.6\\
307 & Galveston County & TX & 48167 & 31.0 & 75.6\\
308 & Sussex County & DE & 10005 & 30.8 & 76.8\\
309 & Caddo County & LA & 22017 & 30.7 & 88.4\\
310 & Charleston County & SC & 45019 & 30.7 & 84.5\\
311 & Early County & GA & 13099 & 30.5 & 85.6\\
312 & Oklahoma County & OK & 40109 & 30.5 & 91.0\\
313 & Rutherford County & TN & 47149 & 30.4 & 85.9\\
314 & St. John the Baptist County & LA & 22095 & 30.1 & 71.9\\
315 & Virginia Beach County & VA & 51810 & 30.0 & 74.8\\
316 & Brazoria County & TX & 48039 & 29.8 & 70.6\\
317 & Manatee County & FL & 12081 & 29.4 & 79.4\\
318 & Livingston County & LA & 22063 & 29.4 & 72.0\\
319 & Kent County & DE & 10001 & 29.4 & 72.0\\
320 & Pinellas County & FL & 12103 & 29.3 & 78.2\\
321 & De Soto County & LA & 22031 & 29.2 & 83.4\\
322 & Lake County & FL & 12069 & 29.1 & 77.2\\
323 & Guilford County & NC & 37081 & 29.0 & 90.2\\
324 & Henry County & GA & 13151 & 29.0 & 81.3\\
325 & Rapides County & LA & 22079 & 29.0 & 75.3\\
326 & Jefferson County & LA & 22051 & 28.8 & 69.7\\
327 & Hillsborough County & FL & 12057 & 28.6 & 78.3\\
328 & Chesapeake County & VA & 51550 & 28.6 & 74.3\\
329 & Lee County & GA & 13177 & 28.2 & 84.9\\
330 & Pasco County & FL & 12101 & 27.5 & 78.5\\
331 & Dougherty County & GA & 13095 & 27.4 & 85.3\\
332 & St. Bernard County & LA & 22087 & 27.3 & 69.6\\
333 & Terrebonne County & LA & 22109 & 27.1 & 70.1\\
334 & Ouachita County & LA & 22073 & 27.1 & 88.9\\
335 & Iberia County & LA & 22045 & 27.0 & 70.6\\
336 & Lubbock County & TX & 48303 & 27.0 & 74.4\\
337 & Brevard County & FL & 12009 & 26.9 & 79.1\\
338 & Houston County & GA & 13153 & 26.3 & 81.1\\
339 & Mitchell County & GA & 13205 & 26.0 & 86.3\\
340 & Calcasieu County & LA & 22019 & 25.4 & 77.5\\
341 & Lee County & AL & 1081 & 25.1 & 80.2\\
342 & Polk County & FL & 12105 & 24.8 & 80.7\\
343 & Volusia County & FL & 12127 & 24.6 & 81.9\\
344 & Carroll County & GA & 13045 & 24.4 & 76.5\\
345 & Beaufort County & SC & 45013 & 24.3 & 77.9\\
346 & York County & SC & 45091 & 24.1 & 80.8\\
347 & Knox County & TN & 47093 & 23.9 & 89.5\\
348 & Jackson County & MS & 28059 & 23.8 & 66.8\\
349 & Chatham County & GA & 13051 & 23.7 & 79.3\\
350 & St. Lucie County & FL & 12111 & 23.5 & 80.3\\
351 & Richland County & SC & 45079 & 23.5 & 84.2\\
352 & Hall County & GA & 13139 & 23.1 & 88.3\\
353 & Clayton County & GA & 13063 & 23.1 & 80.6\\
354 & Douglas County & GA & 13097 & 22.9 & 77.0\\
355 & Greenville County & SC & 45045 & 22.7 & 83.3\\
356 & Duval County & FL & 12031 & 21.9 & 82.8\\
357 & Horry County & SC & 45051 & 21.6 & 81.9\\
358 & Mobile County & AL & 1097 & 21.3 & 70.3\\
359 & Clay County & FL & 12019 & 21.3 & 78.2\\
360 & Muscogee County & GA & 13215 & 21.0 & 74.4\\
361 & Chambers County & AL & 1017 & 20.9 & 82.3\\
362 & Sumter County & GA & 13261 & 20.8 & 80.6\\
363 & Lexington County & SC & 45063 & 19.8 & 83.3\\
364 & Escambia County & FL & 12033 & 19.3 & 73.5\\
365 & Bartow County & GA & 13015 & 19.3 & 77.9\\
366 & Kershaw County & SC & 45055 & 18.8 & 83.2\\
367 & Spartanburg County & SC & 45083 & 17.0 & 84.8\\
368 & Sumter County & SC & 45085 & 16.5 & 79.0\\

\end{longtable}
\end{spacing}

\end{document}